\documentclass[12pt]{article}
\usepackage[margin=1.2in]{geometry}

\usepackage{graphicx,url}

\usepackage[utf8]{inputenc}  
\usepackage{authblk} % for authors' affiliations

\usepackage{amsmath,amsfonts,amssymb}
\usepackage{natbib}
\usepackage{hyperref}
\usepackage{bm} % to use \bm{} for bold greek letters
\usepackage{pdfpages}
\usepackage{array}
\usepackage{MnSymbol} % for \bigoplus, \bigominus
    
\title{Poisson-Tweedie mixed-effects model: a flexible approach for the analysis of longitudinal RNA-seq data}

\author[1]{Mirko Signorelli}
\author[2]{Pietro Spitali}
\author[3]{Roula Tsonaka}

\affil[1]{Department of Biomedical Data Sciences, Leiden University Medical Center (NL)}
\affil[2]{Department of Human Genetics, Leiden University Medical Center (NL)}

\date{}

\begin{document} 

\maketitle

\noindent \textbf{About this article}
\begin{itemize}
\item Please cite this article as: Signorelli, M., Spitali, P., Tsonaka, R. (2020). Poisson–Tweedie mixed-effects model: A flexible approach for the analysis of longitudinal RNA-seq data. \textit{Statistical Modelling}. DOI: 10.1177/1471082X20936017.
\item This document contains the ``accepted'' version of the manuscript. The final (published) version of the article can be freely downloaded (Open Access) from the website of Statistical Modelling, using this link:\\ \href{https://doi.org/10.1177/1471082X20936017}{https://doi.org/10.1177/1471082X20936017}
\end{itemize}

\begin{abstract}
\noindent
We present a new modelling approach for longitudinal count data that is motivated by the increasing availability of longitudinal RNA-sequencing experiments. The distribution of RNA-seq counts typically exhibits overdispersion, zero-inflation and heavy tails; moreover, in longitudinal designs repeated measurements from the same subject are typically (positively) correlated.
We propose a generalized linear mixed model based on the Poisson-Tweedie distribution that can flexibly handle each of the aforementioned features of longitudinal overdispersed counts. We develop a computational approach to accurately evaluate the likelihood of the proposed model and to perform maximum likelihood estimation. Our approach is implemented in the \texttt{R} package \texttt{ptmixed}, which can be freely downloaded from CRAN. We assess the performance of \texttt{ptmixed} on simulated data and we present an application to a dataset with longitudinal RNA-sequencing measurements from healthy and dystrophic mice.
The applicability of the Poisson-Tweedie mixed-effects model is not restricted to longitudinal RNA-seq data, but it extends to any scenario where non-independent measurements of a discrete overdispersed response variable are available.\\


\end{abstract}

\section{Introduction}
\label{sec1}

In the last decade, RNA-sequencing (RNA-seq, \citealt{mortazavi2008}) has progressively replaced microarrays and has become the standard method to measure gene expression. A typical goal of RNA-seq studies is the identification of genes that are differentially expressed across groups of subjects, or are associated with a variable of interest. To this end, a statistical model is postulated to relate the expression levels of a gene to the variable(s) of interests and relevant confounders, and hypothesis testing is performed to assess the significance of the relevant differences or associations. 
While microarrays yielded continuous measurements that could be analysed using models that relied on normality assumptions \citep{smyth2004}, RNA-seq measurents are discrete and require statistical models more suitable for the analysis of count data. 
While the very first attempts to model RNA-seq data relied on the use of the Poisson distribution \citep{marioni2008}, it was soon observed that the distribution of RNA-seq counts is typically characterized by phenomena such as \textit{overdispersion} (the variation of counts is higher than the mean), \textit{zero-inflation} and \textit{heavy tails}, that the Poisson distribution cannot account for. 

To take overdispersion into account, \citet{robinson2010-edgeR} and \citet{love2014} proposed two statistical models based on the negative binomial (NB) distribution, implemented in the \texttt{R} packages \texttt{edgeR} and \texttt{DESeq2}. To deal with small sample sizes, which were a very common feature of the first cross-sectional RNA-seq studies, they proposed to estimate the dispersion parameter of the NB distribution by borrowing information across genes. \cite{esnaola2013} tested the adequacy of the NB distribution on data from a large-scale RNA-seq experiment. They showed that although the NB fitted sufficiently well the empirical distribution of counts for many genes, such distribution was inadequate for a relevant proportion of genes that exhibit strong zero-inflation or heavy tails. Therefore, they proposed a more flexible generalized linear model (GLM) based on the \textit{Poisson-Tweedie} (PT) distribution, implemented in the \texttt{R} packages \texttt{tweeDEseq}. The PT is a three-parameter distribution that encompasses the NB as special case, and differently from the NB it can accommodate different levels of zero-inflation and heavy-tailness for a given amount of overdispersion. A different approach, called limma-voom, was proposed by \cite{law2014}, who combined a linear model for the log-transformed counts with sample-specific precision weights, making it possible to apply to RNA-seq measurements a pipeline originally developed for microarray data.

At the beginning of the diffusion of RNA-seq, the high sequencing costs discouraged the setup of longitudinal RNA-seq studies; for this reason, limma-voom, edgeR, DESeq2 and tweeDEseq were developed to deal with cross-sectional RNA-seq datasets. However, interest in the dynamic evolution of gene expression and declining sequencing costs are nowadays motivating more longitudinal studies. One such example is a mice experiment carried out at the Leiden University Medical Center (LUMC) that aimed to identify biomarkers of disease progression for Duchenne Muscular Dystrophy (DMD). In this experiment, described more in detail in Section \ref{sec4}, blood samples were collected from healthy and dystrophic mice every 6 weeks over a period of 30 weeks, and gene expression in blood was quantified using RNA-seq.

In longitudinal studies, repeated measurements are collected from the \textit{same} subject at several time points. This introduces correlation between the repeated measurements from the same subject, a correlation that is not present in cross-sectional designs where all samples can be assumed to be independent. Nevertheless, often longitudinal RNA-seq data are analysed using popular approaches (mostly edgeR, DESeq2 and limma-voom) for cross-sectional RNA-seq data, either ignoring the longitudinal nature of the data and assuming samples from the same individual to be independent, or adding for each individual a subject-specific indicator as a covariate (fixed effect). 
\cite{cui2016} showed that modelling longitudinal RNA-seq data using such approaches results into increased false positive rates (FPR), or in a loss of power when the FPR is controlled. They argued that longitudinal RNA-seq data should instead be analyzed using generalized linear mixed models (GLMMs, \citealp{mcculloch2008}) and they illustrated their point by comparing those ``improper'' modelling approaches to a Poisson GLMM. They motivated their choice to use the Poisson distribution instead of a more flexible alternative based on frequent convergence problems that they encountered when attempting to fit a negative binomial GLMM with random intercept. Although their choice of a Poisson GLMM simplifies model estimation, it is clear that such a model cannot account for the different combinations of overdispersion, zero-inflation and heavy tails typical of RNA-seq counts. Therefore, the use of GLMMs based on more flexible distributions for count data, such as the NB and PT distributions, should be preferred even if their estimation is more involved.

Recent proposals tailored to the analysis of longitudinal RNA-seq data include MaSigPro \citep{nueda2014}, ShrinkBayes \citep{vandewiel2014}, DyNB \citep{aijo2014} and timeSeq \citep{sun2016}. 
MaSigPro fits a NB GLM that lets the expected value of a gene depend on time polynomials and their interaction with a group indicator; similarly to the methods for cross-sectional RNA-seq, this approach relies on a simple GLM that does not model properly the repeated measurements design. 
ShrinkBayes can fit Poisson, NB, zero-inflated Poisson (ZIP) and zero-inflated NB (ZINB) GLMMs that can accommodate a random intercept. Owing to the possibility to choose the ZIP and the ZINB as distributions, this approach can be employed to model both zero-inflation and the correlation between repeated measurements. However, ShrinkBayes does not allow to fit GLMMs for heavy-tailed distributions. Moreover, the method requires the user to preselect one specific distributional shape that is fitted to all genes, rather than letting the estimation procedure determine it automatically and separately for each gene as in the case of the PT family of models proposed by \cite{esnaola2013}. Finally, often interest may lie in assessing the joint significance of two or more coefficients, for example that of a main effect and of the corresponding interaction term, or in testing a system of linear combinations of the regression coefficients (an example of this type of hypotheses can be found in Section 4). However, the \texttt{R} package  \texttt{ShrinkBayes} allows users to test only hypotheses that involve a single regression coefficient (although with some additional programming one may still be able test more complex hypotheses).
DyNB fits a semiparametric model that combines a NB GLM with Gaussian-processes regression. Finally, timeSeq considers a NB GLMM where time is modelled using splines. Both DyNB and timeSeq are semiparametric and restricted to the NB case, thus they do not attempt to model zero-inflation and heavy tails.

In this paper we present \textit{ptmixed}, a Poisson-Tweedie GLMM that is suitable for the analysis of longitudinal RNA-seq data and that, more in general, can be employed to flexibly model longitudinal counts that are overdispersed, zero-inflated and/or heavy tailed. Owing to the use of the PT distribution, our approach is capable to capture different combinations of zero-inflation and heavy-tails that typically characterize the distribution of overdispersed counts, and it automatically performs model selection among a range of different discrete distributions separately for each gene, similarly to tweeDEseq \citep{esnaola2013}. Moreover, our mixed-effects approach allows to properly model within-subject correlations \citep{cui2016} and to flexibly test both one-dimensional and multi-dimensional statistical hypotheses. To the best of our knowledge, our proposal is the first extension of the PT GLM of \cite{esnaola2013} to the context of random effects models, while a different extension based on generalized estimating equations has been proposed by \cite{bonat2018}. We provide an \texttt{R} implementation of functions for the estimation the proposed model and related hypothesis testing through the \texttt{R} package \texttt{ptmixed} \citep{signorelli-ptmixed}, which can be freely downloaded from the Comprehensive R Archive Network (CRAN). 

The remainder of the article is organized as follows: in Section \ref{sec2} we introduce our model, describe the use of the adaptive Gauss Hermite quadrature rule to approximate its likelihood, present our computational approach to maximum likelihood estimation and discuss the use of Wald and likelihood ratio tests for hypothesis testing. In Section \ref{sec3} we assess the performance of our method using simulated data, and we compare it to several alternative methods that are commonly used in the analysis of longitudinal RNA-seq studies. In Section \ref{sec4} we illustrate an application of ptmixed to the data from our motivating experiment. We conclude by discussing advantages and limitations of our approach in Section \ref{sec5}.

\section{Methods}
\label{sec2}

\subsection{Filtering and normalization of RNA-seq data}
\label{sub:normalization}

The analysis of RNA-seq counts requires careful consideration of two pre-processing steps: filtering and normalization. First, because the expression profiles of lowly expressed genes are typically very noisy and strongly subject to measurement error, those genes are usually filtered out and excluded from the statistical analysis \citep{sha2015}. Filtering allows to focus on those genes for which sufficient information is available and to mitigate the severity of multiple testing corrections. 
Furthermore, gene expression profiles are normalized to remove the effect of systematic technical biases due to the specific technology used to measure gene expression. Hereafter we assume that lowly-expressed genes have been preliminarly filtered out, and that a vector of sample-specific normalization weights has been obtained using one of the available RNA-seq normalization methods, such as for example the \textit{Trimmed Mean of M values} (TMM) method of \cite{robinson2010-tmm}.

\subsection{Poisson-Tweedie generalized linear mixed model}

We consider a longitudinal design in which $j = 1, ..., m_i$ repeated measurements on $g = 1,..., G$ genes are available for $i = 1,..., n$ individuals or subjects. The Poisson-Tweedie mixed-effects model assumes that conditionally on a vector of subject-specific random effects $\textbf{v}_g = (v_{g1}, ..., v_{gn})$, the expression value of gene $g$ measured in the $j$-th sample from individual $i$, $Y_{gij}$, follows a PT distribution with mean $\mu_{gij}$, dispersion $D_g$ and power $a_g$:
\begin{equation}
\begin{gathered}
Y_{gij} \mid v_{gi} \sim \text{PT}(\mu_{gij}, \: D_g, \: a_g) \\
\hspace{1.3cm}
\log \left( \mu_{gij} \right) = x_{ij}^T \beta_g + z_{ij}^T v_{gi} + o_{ij}
\hspace{1.3cm} \\
v_{gi} \stackrel{\text{i.i.d.}}{\sim} N(0, \Sigma_g),
\end{gathered}
\label{pt-glmm}
\end{equation}
where the mean expression profile $\mu_{gij}$ depends on a vector of fixed effects $x_{ij}$, a vector of random effects $z_{ij}$ and on an offset term $o_{ij}$ that corresponds to the logarithm of the sample-specific normalization weights discussed in Section \ref{sub:normalization}. $x_{ij}$ will typically comprise both covariates of interests and relevant confounders. Although in principle it is possible to consider any random effect structure for $z_{ij}$, hereafter we focus on the case in which $z_{ij} = 1$ in \eqref{pt-glmm}, so that $v_{gi} \stackrel{\text{i.i.d.}}{\sim} N(0, \sigma^2_g)$ denotes a subject-specific random intercept.

The PT distribution \citep{el2011}, also known as generalized NB distribution \citep{gerber1991}, is a very flexible three-parameter distribution for discrete data that can be characterized in terms of a mean $\mu > 0$, dispersion $D \geq 1$ and power $a\leq 1$. It encompasses several discrete random variables as special cases, including the Poisson ($a=1, D = 1$), Poisson inverse Gaussian ($a = 0.5$), NB ($a=0$), P\'olia-Aepply ($a = -1$) and Neyman type A ($a \to -\infty$) distributions. \cite{el2011} pointed out that the PT is particularly useful to account for extra zero-inflation or heavy tails compared to the NB distribution.
In Figure 1 of Supplementary File 1 we provide a graphical illustration of the flexibility of the PT, comparing the probability mass function (pmf) of the PT distribution for different values of $D$ and $a$. When $D$ is close to the boundary value 1, the PT pmf is very similar to that of a Poisson distribution irrespective of the value of $a$. For larger values of $D$, the difference between pmfs becomes more noticeable: zero-inflation arises with negative values of $a$, while heavy tails are obtained when $a$ is positive.

\cite{esnaola2013} proposed to exploit the flexibility of the PT in the analysis of cross-sectional RNA-seq data and introduced the PT generalized linear model (GLM), highlighting how the PT power parameter allows to perform automatic model selection from a wide range of different discrete distributions. They showed that for several genes the distribution of RNA-seq counts cannot be assumed to be NB, and that in those cases the use of the PT distribution allows to achieve a better model fit, to improve the estimation of the dispersion parameter and to control the type I error of tests on differences in mean between two groups already with moderate sample sizes. In this article we extend the model of \cite{esnaola2013} by introducing the Poisson-Tweedie GLMM of equation \eqref{pt-glmm}, thus extending the applicability of the PT class of models to the analysis of longitudinal data.

\subsection{Maximum likelihood estimation}
\label{sub:max-lik}

Maximum likelihood (ML) estimation of mixed models is usually carried out by maximizing the marginal likelihood
$$ L(\theta) =  \prod_{i=1}^n \prod_{j=1}^{m_i} f(y_{gij}; \theta) = \prod_{i=1}^n \prod_{j=1}^{m_i} \int_{-\infty}^{\infty} f(y_{gij}|v_{gi}; \theta) f(v_{gi}; \theta) dv_{gi},$$
whose computation requires the integration of the joint likelihood of the response $\bold{y}_g$ and the random effects $\bold{v}_g$ over the distribution of the random effects.
While for the linear mixed model it is possible to perform this integration analytically, for more complex GLMMs computation of the marginal likelihood requires a numeric approximation of the integrals over the random effects (we will discuss this in Section \ref{sub:aghq}).

The marginal likelihood of the PT mixed model with random intercept can be written as
\begin{equation}
L(\beta_g, D_g, a_g, \sigma^2_g | \bold{y}_g, X) =
 \prod_{i=1}^n \int_{-\infty}^{\infty} 
  \frac{1}{\sqrt{ 2 \pi \sigma^2_g} } 
  e^{ - \frac{ v_{gi}^2 }{2 \sigma^2_g} } 
  \prod_{j=1}^{m_i} \int_0^{\infty} \frac{z^{y_{gij}} e^{-z}}{y_{gij}!} f(z) dz \: dv_{gi},
  \label{likelihood}
\end{equation}
where $f(z) = f(z | \beta_g, D_g, a_g)$ denotes the density of the Tweedie distribution. If a more complex random effects structure is specified, then the unidimensional integrals in $dv_{gi}$ in \eqref{likelihood} have to be replaced with multidimensional ones.

A major complication with the computation of \eqref{likelihood} is that because the pmf of the PT distribution does not have a closed-form expression, the computation of the marginal likelihood requires numeric evaluation for both of the integrals in \eqref{likelihood}. As a consequence, ML estimation requires the implementation of a combination of numerical integration and optimization techniques that we illustrate in Sections \ref{sub:aghq} and \ref{sub:mle}.

\subsubsection{Evaluation of the likelihood function}
\label{sub:aghq}

To compute the marginal likelihood in \eqref{likelihood} we first need to evaluate numerically the pmf $\int_0^{\infty} \frac{z^{y_{gij}} e^{-z}}{y_{gij}!} f(z) dz$ of the PT distribution, and then to carry out a numeric approximation of the integrals with respect to the random effects. We base the numeric evaluation of the PT pmf on the recursive formula proposed by \citet{el2011}, resorting in particular to its implementation in the \texttt{R} package \texttt{tweeDEseq} \citep{esnaola2013}. Then, we implement the \textit{adaptive Gauss-Hermite quadrature} (AGHQ) method \citep{pinheiro2006} to approximate the improper integrals associated to the random effects.

To illustrate how the AGHQ method works, we begin by explaining the simpler Gauss-Hermite quadrature (GHQ) method. Let 
$$h(v_{gi}) =  \frac{1}{\sqrt{ 2 \pi \sigma^2_g} } e^{ - \frac{ v_{gi}^2 }{2 \sigma^2_g} } \prod_{j=1}^{m_i} \int_0^{\infty} \frac{z^{y_{gij}} e^{-z}}{y_{gij}!} f(z) dz,$$
so that \eqref{likelihood} can be rewritten as $\prod_{i=1}^n \int_{-\infty}^{\infty} h(v_{gi}) dv_{gi}$. To approximate each integral $\int_{-\infty}^{\infty} h(v_{gi}) dv_{gi}$ with the GHQ method one first needs to rewrite the integrand $h(v_{gi})$ as the product between the kernel $w(v_{gi}) = e^{-v_{gi}^2}$ and a function $q(v_{gi}) = h(v_{gi}) e^{v_{gi}^2}$. Then, the integral of $q(v_{gi})$ with respect to the kernel $w(v_{gi})$ is approximated with a weighted average of the function $q$ evaluated at $k$ predetermined abscissae $v_{gi}^1, v_{gi}^2, ..., v_{gi}^k$:
\begin{equation}
\int_{-\infty}^{\infty} h(v_{gi}) dv_{gi} = \int_{-\infty}^{\infty} w(v_{gi}) q(v_{gi}) dv_{gi} \approx \sum_{r=1}^k w_r q(v_{gi}^r),
\label{eq:agh1}
\end{equation}
where $(w_1, ..., w_k)$ is a vector of known weights \citep{steen1969}. The quality of the approximation in \eqref{eq:agh1}, however, depends on whether the mass of $q(x)$ is concentrated around 0; for this reason, in the AGHQ method the numerical integration is instead performed through an adaptive algorithm whereby one first obtains estimates of the mode $\hat{v}$ of $q(v_{gi})$ and of $\hat{s}^2 = - [ d^2 \log q(\hat{v}) / dv_{gi}^2 ]^{-1}$, and then performs the change of variable $u_{gi} = \frac{v_{gi}-\hat{v}}{\hat{s}\sqrt{2}}$, such that 
\begin{equation}
\int_{-\infty}^{\infty} h(v_{gi}) dv_{gi} \approx \hat{s} \sqrt{2} \sum_{r=1}^k w_r q(\hat{v} + \hat{s} \sqrt{2} u_{gi}^r ).
\end{equation}

When $k = 1$, the AGHQ is equivalent to the \textit{Laplace approximation} (LA), which is often employed to approximate the likelihood of GLMMs. Our choice to employ the AGHQ method is motivated by the fact that approximating the likelihood of GLMMs with the simpler but less accurate LA can result in frequent convergence problems during ML estimation. This problem is particularly frequent for mixed-effects extensions of the Poisson GLMM, such as for example the NB GLMM, and it is commonly encountered when analysing longitudinal RNA-seq counts. For example, \cite{cui2016} attempted to fit NB GLMMs to a dataset with longitudinal RNA-seq measurements by employing \texttt{glmmADMB} \citep{fournier2012}, an \texttt{R} package where the likelihood of GLMMs is approximated using the LA, and reported frequent convergence problems that induced them to fit the simpler Poisson GLMM instead of the NB GLMM. However, such convergence problems can often be solved if the likelihood function is approximated using an AGHQ with a suitable number of quadrature points instead of the LA. Moreover, ML estimates of the variance components obtained with the LA  have been shown to suffer from severe bias, which vanishes if an AGHQ approximation with a sufficient number of quadrature points is used \citep{pinheiro2006}. Overall, although it is a bit more computationally intensive than the LA, the AGHQ produces more precise evaluations of the likelihood, reduces convergence issues, yields more accurate parameter estimates and allows a better type I error control. In Section \ref{sub:agh-lapl} we will provide a demonstration of the gain associated to the use of the AGHQ in place of the LA.

\subsubsection{Likelihood maximization}
\label{sub:mle}

Estimation of the PT mixed model can be performed using the function \texttt{ptmixed()} available in the \texttt{R} package \texttt{ptmixed} (\citealt{signorelli-ptmixed}). This function maximizes the marginal likelihood of the model using the computational approach outlined below, and it allows a certain degree of flexibility in the choice of the number of quadrature points, starting points and maximum number of iterations, as well as of the convergence criterion.

A critical point in the efficiency and success rate of the maximization is the choice of good starting values. As initial values for $\beta$ and $\sigma^2$ we employ the maximum likelihood estimates of those parameters from a negative binomial GLMM or, when its estimation fails, from a Poisson GLMM; for the dispersion parameter $D$ we consider either a method of moments estimate, or the ML estimator of $D$ obtained by fitting a negative binomial GLMM through the function \texttt{nbmixed()}, that estimates the simpler NB GLMM. For the power parameter $a$, we use the method of moment estimator of $a$ for the density of the PT distribution.

Numeric maximization of the likelihood function in \eqref{likelihood} is carried out using the Nelder-Mead (NM, \citealt{nelder1965}) algorithm (as implemented in the function \texttt{optim()}); when the NM method fails, a second attempt is made using the Broyden-Fletcher-Goldfarb-Shanno (BFGS) algorithm. Because NM may require to evaluate the likelihood function several thousands of times, \texttt{ptmixed} offers the possibility to either update the position of the quadrature points at each iteration of the NM algorithm, or after every $k$ iterations. In practice, maximization of the likelihood when the variance of the random intercept is very small may sometimes fail irrespective of the starting values and algorithm used; for this reason, when the starting value for $\sigma^2 < \varepsilon$ we remove the random intercept and fit a PT GLM (we set $\varepsilon = 0.001$ as default in \texttt{ptmixed()}).

\subsection{Inference}
\label{sub:inference}

We base the evaluation of the standard errors of the parameter estimates on the asymptotic distribution of the ML estimator \citep{el2011}, and use the Wald and the likelihood ratio tests for hypotheses testing. If $\hat{\theta}$ is the ML estimate of $\theta = (\beta, D, a, \sigma^2)$, then the asymptotic distribution of $\hat{\theta}$ is multivariate normal with mean $\theta$ and variance $V = \hat{Var}(\hat{\theta}) = I^{-1}(\hat{\theta})$, where $I(\hat{\theta})$ denotes the observed Fisher information evaluated at $\theta = \hat{\theta}$ \citep{mcculloch2008}. %The standard error of the regression coefficient $\beta_k$ can be thus obtained as $SE(\hat{\beta}_k) = \sqrt{V_{kk}}$, where $V_{kk}$ is the element of position $(k, k)$ in $V$. 

Hypotheses that involve linear combinations of regression coefficients can be expressed as $H_0: K^T \beta = K^T b_0$ and tested with the Wald test statistic
$$w = (K^T\hat{\beta}-K^T b_0)^T \left(K^T Var(\hat{\beta}) K \right)^{-1} (K^T\hat{\beta}-K^T b_0),$$
whose asymptotic distribution when $H_0$ is true is $\chi^2_d$, with $d = \text{rank}(K)$. 
Hypotheses that can be expressed in the form $H_0: \theta \in \Theta_0$, where $\Theta_0$ is a subset of the parameter space $\Theta$, can be verified with the the likelihood ratio test (LRT) statistic $t = 2 \log L(\hat{\theta}) - 2 \log L(\hat{\theta}_0)$, where $\hat{\theta}_0 = \text{argmin}_{\theta \in \Theta_0} L(\theta)$. When the null hypothesis does not involve testing values of $\theta$ on the boundary of $\Theta$, the asymptotic distribution of $t$ under $H_0$ is $\chi^2_d$, where $d$ is the number of restrictions implied by $H_0$. If, instead, the hypothesis involves values on the boundary of the parameter space, the asymptotic distribution of $t$ is not $\chi^2$ in general; however, it is possible to show that in certain cases, such as for example when $H_0: \sigma^2 = 0$, the asymptotic distribution of $t$ is a mixture of $\chi^2$ distributions \citep{molenberghs2007}.

\subsection{The \texttt{R} package \texttt{ptmixed}}
\label{sub:software}

\begin{table}[ht]
\caption{Overview of the functions currently available in the \texttt{R} package \texttt{ptmixed} (\citealt{signorelli-ptmixed}, package version 0.5.2).
\label{tab:functions-ptmixed}}
\centering
\vspace{0.5cm}
\begin{tabular}{@{}l|l@{}}\hline
Function & Description\\\hline
\texttt{make.spaghetti} & Creates a trajectory (``spaghetti'') plot to visualize the trajectories\\
&  of a longitudinal outcome of interest\\
\texttt{nbglm} & Estimation of a negative binomial GLM\\
\texttt{nbmixed} & Estimation of a negative binomial GLMM\\
\texttt{pmf} & Computes and plots the empirical pmf of the response variable\\
\texttt{ptglm} & Estimation of a Poisson-Tweedie GLM\\
\texttt{ptmixed} & Estimation of a Poisson-Tweedie GLMM\\
\texttt{ranef} & Computation of the random effects BLUP\\
& for the negative binomial and Poisson-Tweedie GLMMs \\
\texttt{summary.ptglm} & Computation of the standard errors and univariate Wald tests\\
& for the negative binomial and Poisson-Tweedie GLMs \\
\texttt{summary.ptglmm} & Computation of the standard errors and univariate Wald tests\\
& for the negative binomial and Poisson-Tweedie GLMMs \\
\texttt{wald.test} & Computation of the multivariate Wald test\\
\end{tabular}
\end{table}

We have implemented the computational approach for ML estimation outlined in Section \ref{sub:max-lik} in the \texttt{R} package \texttt{ptmixed}, which can be freely downloaded from \texttt{CRAN} \citep{signorelli-ptmixed}. The package comprises functions for the estimation of the Poisson-Tweedie GLMM and of simpler models such as the Poisson-Tweedie GLM and the negative binomial GLM and GLMM. Table \ref{tab:functions-ptmixed} provides an overview of the functions currently available in the package; these functions can be used for data visualization (\texttt{pmf} and \texttt{make.spaghetti}); to fit the PT GLMM (\texttt{ptmixed}), NB GLMM (\texttt{nbmixed}), PT GLM (\texttt{ptglm}) and NB GLM (\texttt{nbglm}); to summarize the results of each model fit (\texttt{summary.ptglm} and \texttt{summary.ptglmm}); to perform hypothesis testing (\texttt{wald.test}); and, for the GLMMs, to obtain the best linear unbiased predictor (BLUP) of the random effects (\texttt{ranef}). Finally, a vignette that illustrates how to use these functions can be found at \url{https://cran.r-project.org/package=ptmixed}

\section{Simulations}
\label{sec3}

In this section we evaluate the performance of ptmixed through several simulations.
We begin by evaluating the accuracy of our approach for ML estimation and hypothesis testing across a range of different distributional shapes (from zero-inflated to heavy-tailed) and sample sizes (Section \ref{sub:ptsim}). 
In Section \ref{sub:agh-lapl} we show the added value of approximating the likelihood function with the AGHQ method by comparing it to the LA.
Finally, in Section \ref{sub:gwidesim} we compare the performance of our method to that of several state of the art approaches for the analysis of RNA-seq data.
R scripts to reproduce the results presented here are available at \url{https://github.com/m-signo/ptmixed}.

\subsection{Accuracy of ML estimates and hypothesis testing}
\label{sub:ptsim}

To assess the accuracy of the ML estimates and of the Wald and likelihood ratio tests computed following the procedures outlined in Section \ref{sub:inference}, we simulated data from the PT mixed model in \eqref{pt-glmm} considering five different values of the power parameter $a$, so as to cover a wide range of distributional shapes: zero-inflated ($a = -5, -1$), NB ($a = 0$), heavy-tailed ($a = 0.5$) and Poisson ($a = 1$). We fixed the dispersion parameter $D = 3$ (except for the case $a = 1$, where $D = 1$) and the number of repeated measurements per subject $m_i = 5 \;\; \forall i \in \{1, ..., n\}$, and progressively increased the number of subjects $n \in \{10, 25, 50, 100, 150\}$. We splitted individuals into two groups, denoted by $d_i \in \{0, 1 \}$, of equal size. We let the mean $\mu_{ij}$ depend on a subject-specific random intercept $v_i$, on group $d_i$ and time $t_i$ in such a way that $\log(\mu_{ij}) = \beta_{0} + v_{i} + \beta_{1} d_i + \beta_{2} t_{ij}$, with $v_i \sim N(0, \sigma^2 = 0.5)$.

\begin{table}
\caption{RMSE of $\hat{\beta}_2$ in simulation A. Values in each cell are estimated using 5000 random replicates.
\label{tab:rmse-time}}
\centering
\vspace{0.5cm}
\begin{tabular}{@{}c|ccccc@{}}\hline
$n$ & a =  -5 & a =  -1 & a =  0 & a =  0.5 & a =  1\\\hline
n = 10 & 0.043 & 0.043 & 0.042 & 0.041 & 0.024 \\ 
  n = 25 & 0.026 & 0.027 & 0.026 & 0.025 & 0.015 \\ 
  n = 50 & 0.018 & 0.019 & 0.018 & 0.018 & 0.010 \\ 
  n = 100 & 0.013 & 0.013 & 0.013 & 0.012 & 0.008 \\ 
  n = 150 & 0.010 & 0.011 & 0.010 & 0.010 & 0.006 \\
  \hline
\end{tabular}
\end{table}

In simulation A we let $\beta = (2.5, 0.3, 0)$ and test whether the time effect is null, i.e. $H_0: \beta_2 = 0$. Table \ref{tab:rmse-time} reports the root mean square error (RMSE) of $\hat{\beta}_2$ for different values of $a$ and $n$. We can observe that the magnitude of the RMSE is comparable across different values of $a$, and that it decreases linearly with $\sqrt{n}$ as expected.
Table \ref{tab:t1e-time} compares the false positive rate (FPR) at $\alpha = 0.05$ between the Wald test and LRT for each of the distributions considered and increasing sample size;  we observe that the two tests exhibit similar performance, producing an FPR close to the nominal type I error ($\alpha = 0.05$) already with small sample sizes.

\begin{table}
\caption{False positive rates of the Wald and likelihood ratio tests for different values of $a$ and $n$ in simulation A ($H_0: \beta_2 = 0$, $\alpha = 0.05$). Values in each cell are estimated using 5000 random replicates.
\label{tab:t1e-time}}
\centering
\vspace{0.5cm}
\begin{tabular}{@{}c|ccccc@{}}\hline
\multicolumn{6}{c}{Wald test}\\
 & a =  -5 & a =  -1 & a =  0 & a =  0.5 & a =  1 \\\hline
n = 10 & 0.058 & 0.053 & 0.061 & 0.060 & 0.036 \\ 
  n = 25 & 0.045 & 0.053 & 0.048 & 0.052 & 0.039 \\ 
  n = 50 & 0.048 & 0.051 & 0.057 & 0.053 & 0.033 \\ 
  n = 100 & 0.051 & 0.053 & 0.051 & 0.054 & 0.042 \\ 
  n = 150 & 0.050 & 0.050 & 0.048 & 0.047 & 0.050 \\ 
	\hline
\multicolumn{6}{c}{Likelihood ratio test}\\ 
 & a =  -5 & a =  -1 & a =  0 & a =  0.5 & a =  1\\\hline
n = 10 & 0.050 & 0.048 & 0.053 & 0.053 & 0.023 \\ 
  n = 25 & 0.043 & 0.051 & 0.047 & 0.050 & 0.026 \\ 
  n = 50 & 0.046 & 0.051 & 0.056 & 0.053 & 0.027 \\ 
  n = 100 & 0.051 & 0.053 & 0.052 & 0.059 & 0.040 \\ 
  n = 150 & 0.049 & 0.053 & 0.051 & 0.054 & 0.046 \\ 
	\hline
\end{tabular}
\end{table}

In simulation B, instead, we let $\beta = (2.5, 0, 0.2)$ and test whether the group coefficient is null, i.e. $H_0: \beta_1 = 0$. Also in this case we see that the RMSE of the parameter of interest (here $\hat{\beta}_1$) is of comparable magnitude across different values of $a$, and that as expected it decreases linearly with $\sqrt{n}$ (Table \ref{tab:rmse-group}).
Assessment of the FPR at $\alpha = 0.05$ (Table \ref{tab:t1e-group}) points out that for very small sample sizes (namely $n = 10$) both the Wald test and LRT are anti-conservative. For larger sample sizes, both methods achieve FPRs much closer to $\alpha = 0.05$.

\begin{table}
\caption{RMSE of $\hat{\beta}_1$ in simulation B. Values in each cell are estimated using 5000 random replicates.
\label{tab:rmse-group}}
\centering
\vspace{0.5cm}
\begin{tabular}{@{}c|ccccc@{}}\hline
$n$ & a =  -5 & a =  -1 & a =  0 & a =  0.5 & a =  1 \\ \hline
n = 10 & 0.456 & 0.474 & 0.459 & 0.473 & 0.448 \\ 
  n = 25 & 0.203 & 0.206 & 0.199 & 0.197 & 0.141 \\ 
  n = 50 & 0.207 & 0.210 & 0.208 & 0.204 & 0.203 \\ 
  n = 100 & 0.147 & 0.146 & 0.146 & 0.145 & 0.143 \\  
  n = 150 & 0.119 & 0.122 & 0.119 & 0.122 & 0.117 \\ 
\hline
\end{tabular}
\end{table}

\begin{table}
\caption{False positive rates of the Wald and likelihood ratio tests for different values of $a$ and $n$ in simulation B ($H_0: \beta_1 = 0$, $\alpha = 0.05$). Values in each cell are estimated using 5000 random replicates.
\label{tab:t1e-group}}
\centering
\vspace{0.5cm}
\begin{tabular}{@{}c|ccccc@{}}\hline
\multicolumn{6}{c}{Wald test}\\
 & a =  -5 & a =  -1 & a =  0 & a =  0.5 & a =  1 \\\hline
n = 10 & 0.110 & 0.129 & 0.117 & 0.125 & 0.116 \\ 
  n = 25 & 0.064 & 0.063 & 0.057 & 0.059 & 0.048 \\ 
  n = 50 & 0.056 & 0.058 & 0.060 & 0.056 & 0.062 \\ 
  n = 100 & 0.060 & 0.056 & 0.055 & 0.055 & 0.058 \\ 
   n = 150 & 0.049 & 0.056 & 0.050 & 0.062 & 0.051 \\ 
\hline
\multicolumn{6}{c}{Likelihood ratio test}\\
 & a =  -5 & a =  -1 & a =  0 & a =  0.5 & a =  1 \\\hline
n = 10 & 0.080 & 0.093 & 0.085 & 0.092 & 0.057 \\ 
  n = 25 & 0.058 & 0.060 & 0.054 & 0.056 & 0.031 \\ 
  n = 50 & 0.051 & 0.054 & 0.055 & 0.051 & 0.042 \\ 
  n = 100 & 0.056 & 0.053 & 0.052 & 0.054 & 0.043 \\
  n = 150 & 0.048 & 0.054 & 0.049 & 0.061 & 0.039 \\ 
\hline
\end{tabular}
\end{table}

\subsection{Comparison between the AGHQ and Laplace approximations of the likelihood}
\label{sub:agh-lapl}

To quantify the improvement associated to the use of the AHGQ in the approximation of the likelihood of our model, we illustrate a comparison of its performance to that of the LA, focusing on the case with $a = -1$ and $n = 100$ considered in simulation B (for the description of the settings of this simulation, see Section \ref{sub:ptsim}). Table \ref{tab:lapl1} compares the mean of the ML estimates obtained when using the LA or an AGHQ with 5 quadrature points. The results show that even for $n = 100$ the LA can yield rather biased parameter estimates, in particular for $D$, $a$ and $\sigma^2$. Use of the AGHQ makes it possible to obtain more accurate parameter estimates for all parameters.

\begin{table}[h]
\caption{Evaluation of the bias of the ML estimates obtained using the LA and the AGHQ. Values in each cell are estimated using 5000 random replicates.
\label{tab:lapl1}}
\centering
\vspace{0.5cm}
\begin{tabular}{@{}c|ccc@{}}\hline
 & true value & Laplace & AGHQ \\ 
  \hline
$\beta_0$ & 2.500 & 2.483 & 2.493 \\ 
$\beta_1$ & 0.000 & 0.003 & 0.001 \\ 
$\beta_2$ & 0.200 & 0.202 & 0.199 \\ 
$D$ & 3.000 & 2.612 & 3.108 \\ 
$a$ & -1.000 & -2.520 & -0.653 \\ 
$\sigma^2$ & 0.500 & 0.435 & 0.471 \\ \hline
\end{tabular}
\end{table}

Importantly, use of the AGHQ produced considerably higher convergence rates (87.5\% versus the 65.2\% of the LA) using the default starting values computed by \texttt{ptmixed} (thus reducing the risk that one may need to try different starting points to ensure convergence). Furthermore, the AGHQ reduced the proportion of false positives (5.6\% versus 7\% for the LA, $\alpha = 0.05$) enabling a better type I error control. All these performance improvements are due to the fact that the use of more quadrature points allowed a better approximation of the likelihood function, and they required on average a 12\% increase in computing time with respect to the LA (Table \ref{tab:lapl2}).

\begin{table}[h]
\caption{Convergence rate, estimated FPR and mean computing time using the LA and the AGHQ. Values in each cell are estimated using 5000 random replicates.
\label{tab:lapl2}}
\centering
\vspace{0.5cm}
\begin{tabular}{@{}c|cc@{}}\hline
& Laplace & AGHQ\\\hline
Convergence & 65.2\% & 87.5\% \\
False positive rate ($\alpha = 0.05$) & 0.070 & 0.056 \\
Mean computing time (minutes) & 10.3 & 11.5 \\\hline
\end{tabular}
\end{table}

\subsection{Comparison with alternative modelling approaches}
\label{sub:gwidesim}

In this Section we describe the results of two simulation studies (C and D) designed to compare \texttt{ptmixed} to several state of the art approaches that are typically employed in the analysis of longitudinal RNA-seq data. Contrary to Section \ref{sub:ptsim}, where we assessed the performance of our method in parameter estimation and testing for a fixed $\theta = (\beta, D, a, \sigma^2)$, here we evaluate the overall performance of each method on simulated datasets comprising $G = 500$ genes each (i.e., each gene has a different $\theta_g$); the reason for this is that some of the methods included in the comparison (DESeq2, edgeR and limma-voom) carry out parameter estimation by borrowing information across different genes, and therefore they require information from all genes to estimate and test the parameters associated to each individual gene.

We simulate datasets comprising gene expression values of $G = 500$ genes for a sample of $n$ subjects who are splitted into two groups of equal size. We assume that $m_i = 5$ repeated measurements are available for each subject. The data are generated from model \eqref{pt-glmm} assuming that
$$\log(\mu_{gij}) = \beta_{g0} + \beta_{g1} d_i + \beta_{g2} t_{ij} + o_{ij} + v_{gi},$$
where $d_i \in \{0, 1\}$ is the group indicator, $t_{ij} \in \{ 0, 1, ..., 4\}$ represents time, $o_{ij} \sim N(0, 0.5^2)$ is an offset term whose purpose is to mimic the presence of sample-specific normalization weights, $v_{gi} \sim N(0, \sigma^2_g)$ with $\sigma^2_g \sim U(0.2,0.8)$ is a random intercept, and the dispersion parameter $D$ follows a translated gamma distribution, $D-1 \sim \Gamma(\alpha = 2, \beta = 1)$.
Consistently with the results of \citet{esnaola2013}, who showed that the hypothesis that the distribution of RNA-seq counts is NB cannot be rejected for most genes, but that at the same time the distribution is not NB for a relevant proportion of genes, we let the distribution of $Y_{gij}$ to be NB ($a_g = 0$) for 300 genes, zero-inflated ($a_g \sim U(-10, -1)$) for 100 genes and heavy-tailed ($a_g \sim U(0.3, 0.7)$) for the remaining 100 genes.

\begin{table}
\caption{List of methods compared in simulations C and D.
\label{tab:list-methods}}
\centering
\vspace{0.5cm}
\begin{tabular}{@{}c|ccc@{}}\hline
Method & Model type & Assumption about & Reference\\
acronym& & repeated measurements &\\\hline
DESeq2 & NB GLM & Independence & \cite{love2014}\\
DESeq2+FE & NB GLM & Fixed effects & \cite{love2014}\\
edgeR & NB GLM & Independence & \cite{robinson2010-edgeR}\\
edgeR+FE & NB GLM & Fixed effects & \cite{robinson2010-edgeR}\\
limma-voom & Gaussian LM & Independence & \cite{law2014}\\
limma-voom+FE & Gaussian LM & Fixed effects & \cite{law2014}\\
MaSigPro & NB GLM & Independence & \cite{nueda2014} \\
ptmixed & PT GLMM & Random effects & \cite{signorelli-ptmixed}\\
tweeDEseq & PT GLM & Independence & \cite{esnaola2013}\\
ShrinkBayes & ZINB GLMM & Random effects & \cite{vandewiel2014}\\\hline
\end{tabular}
\end{table}

We have included in this comparison approaches that (i) treat repeated measurements from the same individual as independent, (ii) include the subject identifiers as fixed effects into the model, and (iii) use random effects to model the correlation induced by the repeated measurements design. An overview of the methods considered is given in Table \ref{tab:list-methods}, where we summarize the type of model used and the way in which they handle the repeated measurements design. All methods were fitted specifying as fixed effects the main effect of group and time, with the exception of MaSigPro where also an interaction term was included (in the \texttt{R} package \texttt{MaSigPro} the interaction term is automatically included and cannot be suppressed). As concerns hypothesis testing, for ptmixed we assessed the performance of both the Wald test and the LRT introduced in Section \ref{sub:inference}.

In simulation $C$ we compare the performance of each method in testing if a gene is up- or down-regulated over time, i.e., $\beta_{g2} = 0$. We let $\beta_{g0} \sim N(3, 0.5)$, $\beta_{g1} \sim N(0, 0.3)$, and we assume that 400 genes are not dysregulated over time (i.e., $\beta_{g2} = 0$), 50 genes are up-regulated with $\beta_{g2} \sim U(0.1, 0.5)$ and 50 genes downregulated with $\beta_{g2} \sim U(-0.5, -0.1)$.
In simulation $D$, instead, we check the performance of each method in testing if a gene is differentially expressed (DE) between the two groups, i.e., $\beta_{g1} = 0$. We let $\beta_{g0} \sim N(3, 0.5)$, $\beta_{g2} \sim U(-0.1, 0.1)$, and we assume that 400 genes are not DE (i.e., $\beta_{g1} = 0$), 50 genes are up-regulated with $\beta_{g1} \sim U(0.5, 1)$ and 50 genes downregulated with $\beta_{g1} \sim U(-1, -0.5)$.

First, we assess the precision of the parameter estimates obtained by the different methods by comparing their RMSEs. We report the RMSE for all methods with the exception of MaSigPro, for which the computation was not possible because this method reports parameter estimates only for the significant genes, and not for all genes. Then, we evaluate the extent to which each method controls the type I error by computing the False Positive Rate (FPR), which is the proportion of truly non-DE genes that are declared as DE by the test at a fixed significance level $\alpha$. Furthermore, for the methods whose FPR appears to be under control we compare the True Positive Rate (TPR), which is the proportion of truly DE genes that are correctly declared as DE by the test at level $\alpha$. Finally, we also present a comparison of the computing times of each method. The results of all methods were corrected for multiple testing. We employed the Benjamini-Hochberg \citep{benjamini1995} correction for all methods list in Table \ref{tab:list-methods}, with the exception of ShrinkBayes, for which we employed the Bayesian FDR suggested in \cite{vandewiel2014}. Results for each simulation and sample size are based on 50 random replications (i.e., 50 datasets comprising 500 genes each).

\begin{table}
\caption{RMSE of $\hat{\beta}_2$ in simulation C.
\label{tab:rmse-simC}}
\centering
\vspace{0.5cm}
\begin{tabular}{@{}c|ccccc@{}}\hline
 Method & n = 10 & n = 20 & n = 40 \\ 
  \hline
DESeq2 & 0.117 & 0.106 & 0.106 \\ 
  DESeq2+FE & 0.124 & 0.114 & 0.112 \\ 
  edgeR & 0.101 & 0.082 & 0.073 \\ 
  edgeR+FE & 0.106 & 0.088 & 0.078 \\ 
  limma-voom & 0.126 & 0.106 & 0.093 \\ 
  limma-voom+FE & 0.110 & 0.093 & 0.082 \\ 
  ptmixed & 0.120 & 0.051 & 0.039 \\ 
  ShrinkBayes & 0.062 & 0.040 & 0.026 \\ 
  tweeDEseq & 0.066 & 0.054 & 0.049 \\ 
\hline
\end{tabular}
\end{table}

\begin{table}
\caption{RMSE of $\hat{\beta}_1$ in simulation D.
\label{tab:rmse-simD}}
\centering
\vspace{0.5cm}
\begin{tabular}{@{}c|ccccc@{}}\hline
 Method & n = 10 & n = 20 & n = 40 \\ 
  \hline
DESeq2 & 0.910 & 0.743 & 0.648 \\ 
  DESeq2+FE & 4.406 & 4.419 & 4.424 \\ 
  edgeR & 0.756 & 0.549 & 0.408 \\ 
  edgeR+FE & 4.224 & 4.225 & 4.230 \\ 
  limma-voom & 0.789 & 0.570 & 0.428 \\ 
  limma-voom+FE & 23.076 & 23.091 & 23.077 \\ 
  ptmixed & 0.066 & 0.049 & 0.035 \\ 
  ShrinkBayes & 0.347 & 0.247 & 0.180 \\ 
  tweeDEseq & 0.412 & 0.290 & 0.210 \\ 
\hline
\end{tabular}
\end{table}

Table \ref{tab:rmse-simC} compares $RMSE(\hat{\beta}_{g2})$ in simulation $C$, where interest lies in the estimation and testing of the time effect $\beta_{g2}$, which is a within-subjects effect. In this simulation, the method with the lowest RMSEs is ShrinkBayes, closely followed by tweeDEseq when $n = 10$ and $n = 20$, and by ptmixed when $n = 20$ and $n = 40$. The remaining six methods are associated to larger RMSEs across all considered sample sizes.
Table \ref{tab:rmse-simD} compares $RMSE(\hat{\beta}_{g1})$ in simulation $D$, where interest lies in the estimation and testing of the group effect $\beta_{g1}$, which is a between-subjects effect. Here, ptmixed is by far the method with the lowest RMSEs. The RMSEs of ShrinkBayes, tweeDEseq, edgeR, limma and DEseq2 are between 5 and 20 times higher than those of ptmixed, whereas the three approaches that use subject-specific fixed effects have extremely high RMSEs that do not even decrease with $n$. This last result shows that attempts to model longitudinal data using fixed effects can yield extremely inaccurate parameter estimates of between-subjects effects. 

\begin{table}
\caption{False Positive Rate in simulation C ($H_0: \beta_{g2} = 0,\: \alpha = 0.05$).
\label{tab:fpr-simC}}
\centering
\vspace{0.5cm}
\begin{tabular}{@{}c|ccccc@{}}\hline
 Method & n = 10 & n = 20 & n = 40 \\ 
  \hline
DESeq2 & 0.001 & 0.004 & 0.048 \\ 
  DESeq2+FE & 0.052 & 0.179 & 0.471 \\ 
  edgeR & 0.000 & 0.000 & 0.000 \\ 
  edgeR+FE & 0.005 & 0.005 & 0.005 \\ 
  limma-voom & 0.000 & 0.000 & 0.000 \\ 
  limma-voom+FE & 0.008 & 0.009 & 0.011 \\ 
  MaSigPro & 0.000 & 0.000 & 0.000 \\ 
  ptmixed, LRT & 0.023 & 0.030 & 0.044 \\ 
  ptmixed, Wald test & 0.029 & 0.033 & 0.040 \\ 
  ShrinkBayes & 0.032 & 0.031 & 0.025 \\ 
  tweeDEseq & 0.005 & 0.005 & 0.004 \\    
  \hline
\end{tabular}
\end{table}

\begin{table}
\caption{True Positive Rate in simulation C ($H_0: \beta_{g2} = 0,\: \alpha = 0.05$) for the methods that control the type I error (FPR $\leq \alpha$ in Table \ref{tab:fpr-simC}). Asterisks (*) denote methods that did not control the type I error (see Table \ref{tab:fpr-simC}).
\label{tab:tpr-simC}}
\centering
\vspace{0.5cm}
\begin{tabular}{@{}c|ccccc@{}}\hline
 Method & n = 10 & n = 20 & n = 40 \\ 
  \hline
DESeq2 & 0.693 & 0.840 & 0.926 \\ 
  DESeq2+FE & 0.884 & * & * \\ 
  edgeR & 0.634 & 0.815 & 0.937 \\ 
  edgeR+FE & 0.847 & 0.947 & 0.985 \\ 
  limma-voom & 0.622 & 0.807 & 0.927 \\ 
  limma-voom+FE & 0.853 & 0.947 & 0.985 \\ 
  MaSigPro & 0.025 & 0.009 & 0.002 \\ 
  ptmixed, LRT & 0.750 & 0.893 & 0.968 \\ 
  ptmixed, Wald test & 0.769 & 0.895 & 0.968 \\ 
  ShrinkBayes & 0.663 & 0.876 & 0.961 \\ 
  tweeDEseq & 0.736 & 0.860 & 0.951 \\ 
   \hline
\end{tabular}
\end{table}

Table \ref{tab:fpr-simC} reports the estimated FPRs in simulation C. Overall, we can observe that in this case almost all methods control the type I error, achieving an FPR $< 5\%$ for all sample sizes. The only exception to this is DESeq2+FE. It can also be noted that the FPRs of the two mixed-effects modelling approaches, ptmixed (both with the Wald test and the LRT) and ShrinkBayes, are much closer to the prespecified $\alpha$ than those of all other methods, which appear to be more conservative.

The estimated TPRs for the methods that control the type I error in simulation C are shown in Table \ref{tab:tpr-simC}. In general, all methods achieve comparable TPRs, with the exception of MaSigPro whose TPR is almost 0 for all $n$. The methods with the highest TPR when $n = 10$ are the three fixed-effects approaches, whereas for $n = 20$ and $n = 40$ edgeR+FE and limma-voom+FE are the most powerful, followed by ptmixed, ShrinkBayes and tweeDEseq.

\begin{table}
\caption{False Positive Rate in simulation D ($H_0: \beta_{g1} = 0,\: \alpha = 0.05$).
\label{tab:fpr-simD}}
\centering
\vspace{0.5cm}
\begin{tabular}{@{}c|ccccc@{}}\hline
 Method & n = 10 & n = 20 & n = 40 \\ 
  \hline
DESeq2 & 0.286 & 0.289 & 0.323 \\ 
  DESeq2+FE & 0.977 & 0.978 & 0.978 \\ 
  edgeR & 0.234 & 0.234 & 0.238 \\ 
  edgeR+FE & 0.977 & 0.979 & 0.977 \\ 
  limma-voom & 0.194 & 0.184 & 0.190 \\ 
  limma-voom+FE & 1.000 & 1.000 & 1.000 \\ 
  MaSigPro & 0.000 & 0.000 & 0.000 \\ 
  ptmixed, LRT & 0.003 & 0.005 & 0.012 \\ 
  ptmixed, Wald test & 0.031 & 0.012 & 0.012 \\ 
  ShrinkBayes & 0.120 & 0.035 & 0.022 \\ 
  tweeDEseq & 0.342 & 0.316 & 0.307 \\ 
  \hline
\end{tabular}
\end{table}

\begin{table}
\caption{True Positive Rate in simulation D ($H_0: \beta_{g1} = 0,\: \alpha = 0.05$) for the methods that control the type I error (FPR $\leq \alpha$ in Table \ref{tab:fpr-simD}). Asterisks (*) denote methods that did not control the type I error (see Table \ref{tab:fpr-simD}).
\label{tab:tpr-simD}}
\centering
\vspace{0.5cm}
\begin{tabular}{@{}c|ccccc@{}}\hline
 Method & n = 10 & n = 20 & n = 40 \\ 
  \hline
DESeq2 & * & * & * \\ 
  DESeq2+FE & * & * & * \\ 
  edgeR & * & * & * \\ 
  edgeR+FE & * & * & * \\  
  limma-voom & * & * & * \\ 
  limma-voom+FE & * & * & * \\  
  MaSigPro & 0.001 & 0.000 & 0.000 \\ 
  ptmixed, LRT & 0.040 & 0.228 & 0.607 \\ 
  ptmixed, Wald test & 0.206 & 0.336 & 0.630 \\ 
  ShrinkBayes & * & 0.363 & 0.646 \\ 
  tweeDEseq & * & * & * \\ 
   \hline
\end{tabular}
\end{table}

The estimated FPRs in simulation D are presented in Table \ref{tab:fpr-simD}. In this case, ptmixed and MaSigPro control the type I error across all sample sizes, and ShrinkBayes does so for $n = 20$ and $n = 40$. 
All other methods fail to control the type I error, producing much higher FPRs: the approaches that assume independence (DESeq2, edgeR, limma-voom and tweeDEseq) yield FPRs that range from 0.184 to 0.342, whereas the approaches that employ fixed effects (DESeq2+FE, edgeR+FE and limma-voom+FE) have FPRs close to 1.

The estimated TPRs for the methods that control the type I error in simulation D are shown in Table \ref{tab:tpr-simD}. Once again, MaSigPro has TPRs close to 0, giving an indication that the method has little power to reject $H_0$ both in simulation C and simulation D. As concerns ptmixed, the Wald test appears to be a bit more powerful than the LRT, especially when $n = 10$ and $n = 20$. Finally, in the two cases in which it controls the type I error ($n = 20$ and $n = 40$), ShrinkBayes achieves a TPR similar to that of ptmixed's Wald test.

In Tables 1 and 2 of Supplementary File 1 we report the computing time of each method in simulations C and D. Overall, estimation of the GLMM approaches, ptmixed and ShrinkBayes, was considerably slower than that of the other approaches that fit simpler GLMs. For most methods, computing time increased linearly with $n$; this did not hold for the fixed-effects approaches, where computing time increased more than linearly with $n$, and for ShrinkBayes, which took more time to convergence when $n = 10$ than when $n = 20$ or $n = 40$.

\section{Characterizing the dynamic evolution of Duchenne Muscular Dystrophy in mouse}
\label{sec4}

In this section we illustrate how the ptmixed can be employed to analyse longitudinal RNA-seq data. We consider a longitudinal RNA-seq experiment performed at the LUMC (Leiden, NL) whose aim was to characterize the dynamic evolution of gene expression in \textit{mdx} mice. The \textit{mdx} mouse is the most used mouse model for DMD, an X-linked recessive genetic disease caused by protein truncating mutations in the \textit{DMD} gene that encodes dystrophin. The progression of DMD is characterised by a process of muscle degeneration and regeneration whose consequences are muscle instability and the replacement of muscular tissue with fibrotic and fatty tissues. Eventually, this process leads to loss of muscle function, loss of ambulation and premature death. To date no effective treatment for DMD has been found; one of the reason for drug failure during clinical trials is the use functional tests with high inter- and intra-individual variability, resulting in underpowered studies. The possibility to quantify disease progression non-invasively with objective biomarkers is highly needed. In this experiment, repeated blood measurements were taken in order to obtain estimates of the evolution of blood-based gene expression over time.

The experiment involved five different groups of mice, but for simplicity of illustration here we focus on the comparison of the two most interesting groups: wild-type (\textit{wt}), comprising 5 healthy mice, and \textit{mdx}, comprising 3 dystrophic mice that shared the same genetic background of \textit{wt} mice but carried a nonsense mutation in exon 23 that led to lack of dystrophin. Mice were kept under observation for 30 weeks, during which blood samples were collected from each mouse every 6 weeks starting from 6 weeks of age. Transcriptomic expression in these samples was quantified using RNA-sequencing (GEO accession number GSE132741). We focus our attention on those genes that belong to 5 pathways, listed in Table \ref{tab:pathways}, that are relevant in the DMD pathophysiology. We filter out lowly expressed genes by excluding genes that have less than 1 count per million in more than half of the samples; after filtering, 379 genes are retained in the analysis.

\begin{table}[h]
\caption{List of pathways considered in our analysis. The pathways were downloaded form WikiPathways \citep{slenter2017}.
\label{tab:pathways}}
\centering
\vspace{0.5cm}
\begin{tabular}{@{}lll@{}}\hline
Pathway & Description & Number of genes\\ \hline
WP113 & TGF$\beta$ signaling sathway & 48\\
WP246 & TNF$\alpha$ NF-kB signaling pathway & 175\\
WP387 & IL-6 signaling pathway & 29\\
WP3675 & Factors and pathways affecting insulin-like & 96\\
& growth factor (IGF1)-Akt signaling &\\
WP85 & Focal adhesion & 168 \\\hline
\end{tabular}
\end{table}

We let the expression value for gene $g$ from mouse $i$ at the $j$-th time point follow a PT GLMM where 
$ Y_{gij} \mid v_{gi} \sim \text{PT}(\mu_{gij}, \: D_g, \: a_g) $ and 
\begin{equation}
\log(\mu_{gij}) = \beta_{g0} + \beta_{g1} d_i + \beta_{g2} t_{ij} + \beta_{g3} d_i t_{ij} + o_{ij} + v_{gi},
\label{eq:application}
\end{equation}
where $d_i$ is a dummy variable which is 0 for \textit{wt} and 1 for \textit{mdx} mice, $t_{ij} \in \{0, 1, 2, 3, 4\}$ denotes time (0 = week 6, 4 = week 30) and $o_{ij}$ is an offset term equal to the logarithm of the effective library sizes computed using the TMM normalization method \citep{robinson2010-tmm}.

We estimated model \ref{eq:application} using an AGHQ with 10 quadrature points and considered up to two different starting values (\texttt{ptmixed} default values and, when necessary, ML estimates of the negative binomial GLMM) for the optimization. ML estimation was successful for 315 genes; for the remaining 64 genes, for which estimation of model \ref{eq:application} failed, we attempted to estimate the simpler NB mixed model by fixing $a_g = 0$; estimation of such model was successful for 63 of those genes. We do not consider further the remaining gene, for which also estimation of the NB GLMM failed.

\begin{figure}
\centering
\includegraphics[scale = 0.6]{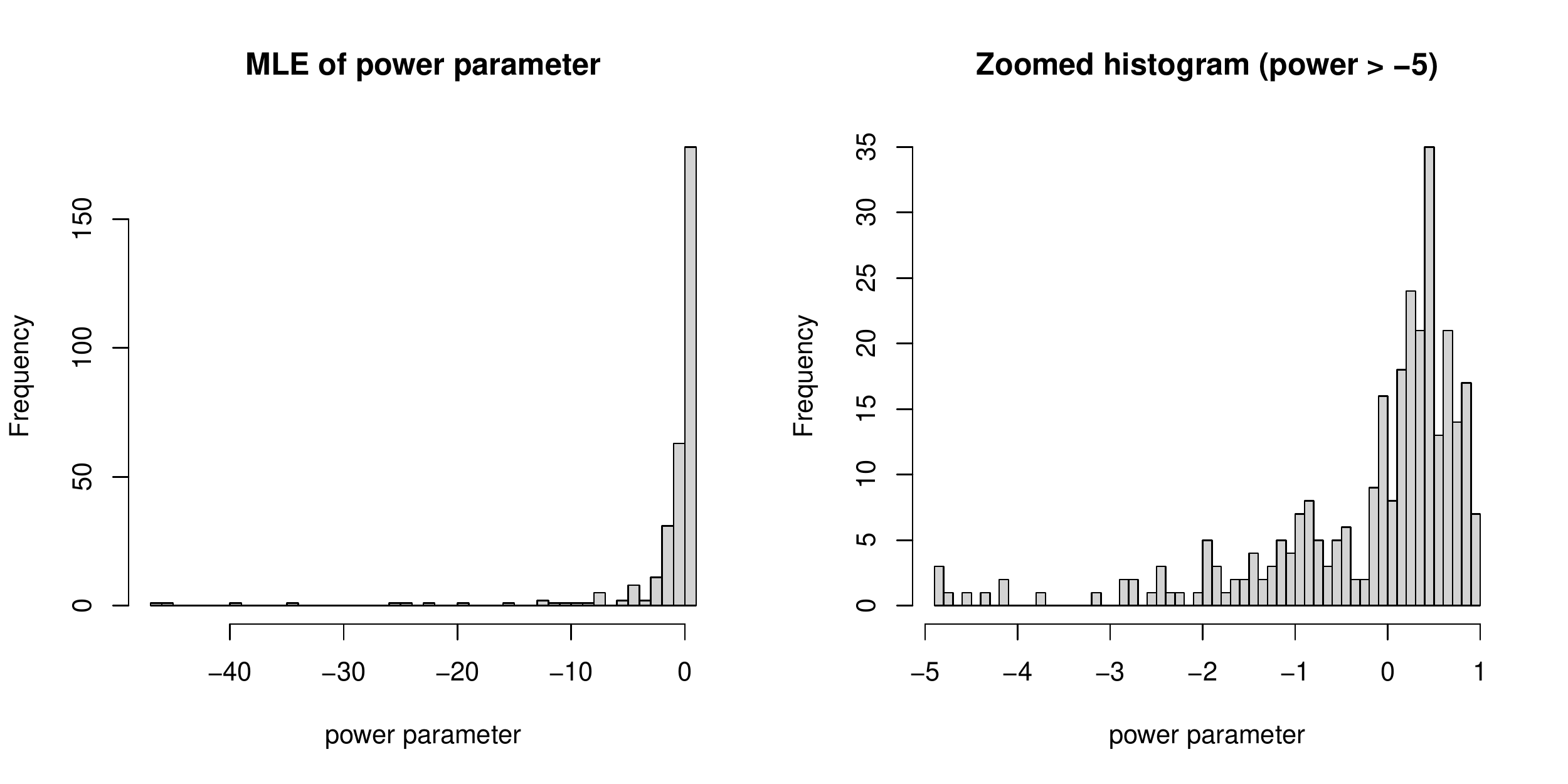}
\caption{Histograms showing the distribution of the ML estimates of the power parameter. The left histogram shows the distribution of the ML estimates over the full range of $\hat{a}_g$, whereas the right histogram zooms on the interval $[-5, 1]$, where most of the distribution is concentrated.}
\label{fig:mle-power}
\end{figure}

Figure \ref{fig:mle-power} shows the distribution of the ML estimates of the power parameter, $\hat{a}_g$, for the 315 genes for which estimation of the PT GLMM was successful. For 74 genes $\hat{a}_g < -1$, for 62 genes $\hat{a}_g \in [-1, 0)$, for 106 genes $\hat{a}_g \in [0, 0.5)$, and for the remaining 72 genes $\hat{a}_g \in [0.5, 1)$. The spread of $\hat{a}_g$ over a range of values far from 0 is an indication of the fact that for several genes the NB distribution (that is, a PT with $a = 0$) does not provide the best fit to the data, in line with the observations made by \cite{esnaola2013} about the distribution of cross-sectional RNA-seq data.

With the aim of identifying genes whose progression is different between \textit{mdx} and \textit{wt} mice, we tested the null hypothesis $H_{01}: \beta_{g1} = \beta_{g3} = 0$ of no difference in the average expression level of gene $g$ between cases and controls at any time point. p-values were adjusted for multiple testing using the Benjamini-Hockberg method \citep{benjamini1995}. Detailed results of the test are provided in Supplementary File 2. In short, we found evidence of differential expression (FDR-corrected p-value $< 0.05$) for 120 genes; for these genes, we furthermore tested whether there was sufficient evidence that the intercept or the slope are different between groups, $H_{02}: \beta_{g1} = 0$ and $H_{03}: \beta_{g3} = 0$ respectively. These tests allowed to further identify 38 genes with a significant intercept (difference between \textit{wt} and \textit{mdx} at week 6) and 38 genes where the progression (slope) is different between groups. 

\begin{figure}
\centering
\includegraphics[scale = 0.4, page = 1]{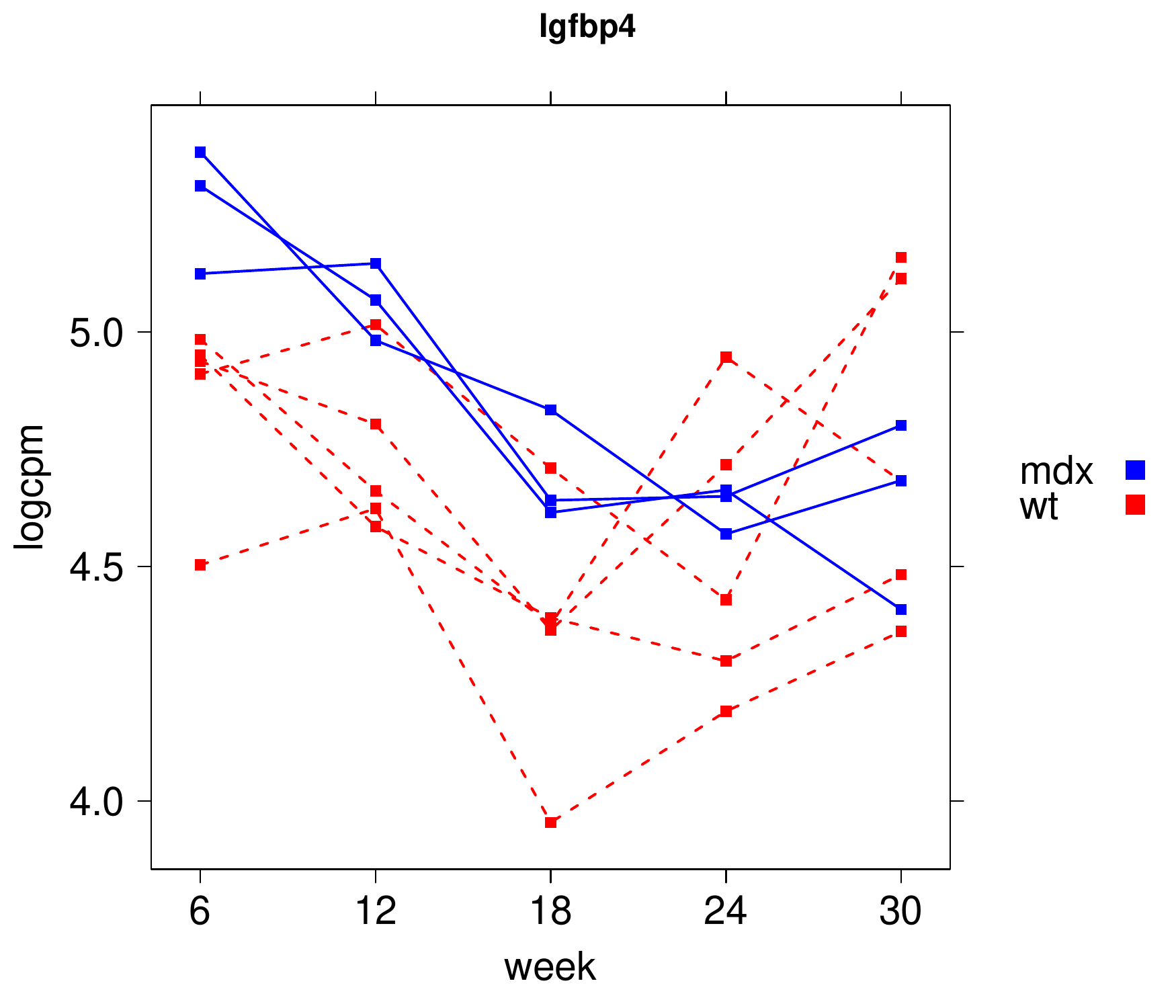}
\includegraphics[scale = 0.4, page = 2]{Fig2.pdf}
\includegraphics[scale = 0.4, page = 3]{Fig2.pdf}
\includegraphics[scale = 0.4, page = 4]{Fig2.pdf}
\caption{Plots comparing the trajectories of selected genes in \textit{mdx} (blue solid line) and \textit{wt} (red dashed line) mice.}
\label{fig:appl-traj}
\end{figure}

In Figure \ref{fig:appl-traj} we show the trajectories of four of the significant genes, \textit{Igfbp4}, \textit{Jak1}, \textit{Stat1} and \textit{Tgif1}. \textit{Igfbp4} is one of the insulin-like growth factor binding proteins, it binds to both IGF-1 and IGF-2 affecting the half-life of IGFs in circulation and their interaction with cell receptors. IGF-1 has been linked to muscle regeneration and multiple papers show a beneficial effect of IGF-1 in DMD \citep{forcina2019}. IGF binding proteins such as IGFBP3 and IGFBP5 have also been described to be reduced in DMD patients serum compared to healthy controls \citep{hathout2019}. In this work we observe a significant elevation of \textit{Igfbp4} in \textit{mdx} mice, which is particularly apparent at weeks 6 and 12 when intensive muscle degeneration and regeneration is ongoing. 
\textit{Jak1} and \textit{Stat1} are part of the JAK-STAT pathway; inspection of the trajectory plots in Figure \ref{fig:appl-traj} shows that the two genes have a similar pattern of expression, with an elevation in \textit{mdx} mice at week 6 followed by a steep decrease at the later time points. It has been described how the JAK-STAT pathway contributes to muscle repair promoting regeneration in DMD \citep{moresi2019}. The observed behaviour of \textit{Jak1} and \textit{Stat1} in \textit{mdx} mice is particularly interesting for two reasons: the intensive muscle regeneration ongoing at 6 weeks of age in \textit{mdx} mice, and the previous finding that \textit{Stat1} is activated in \textit{mdx} mice diaphragm at 6 weeks of age compared to healthy mice \citep{dogra2008}. 
Finally, for \textit{Tgif1} we can observe a downregulation in \textit{mdx} mice at 6 and 12 weeks; later on, the expression of the gene in \textit{mdx} mice increases towards the levels observed in \textit{wt} mice. \textit{Tgif1} is part of the TGF$\beta$ pathway, a well known pathway in the DMD pathophysiology with genetic modifiers mapping to this pathway \citep{flanigan2013} and a number of reports involving TGF$\beta$ in DMD \citep{ceco2013}. 

\begin{figure}
\centering
\includegraphics[scale = 0.5]{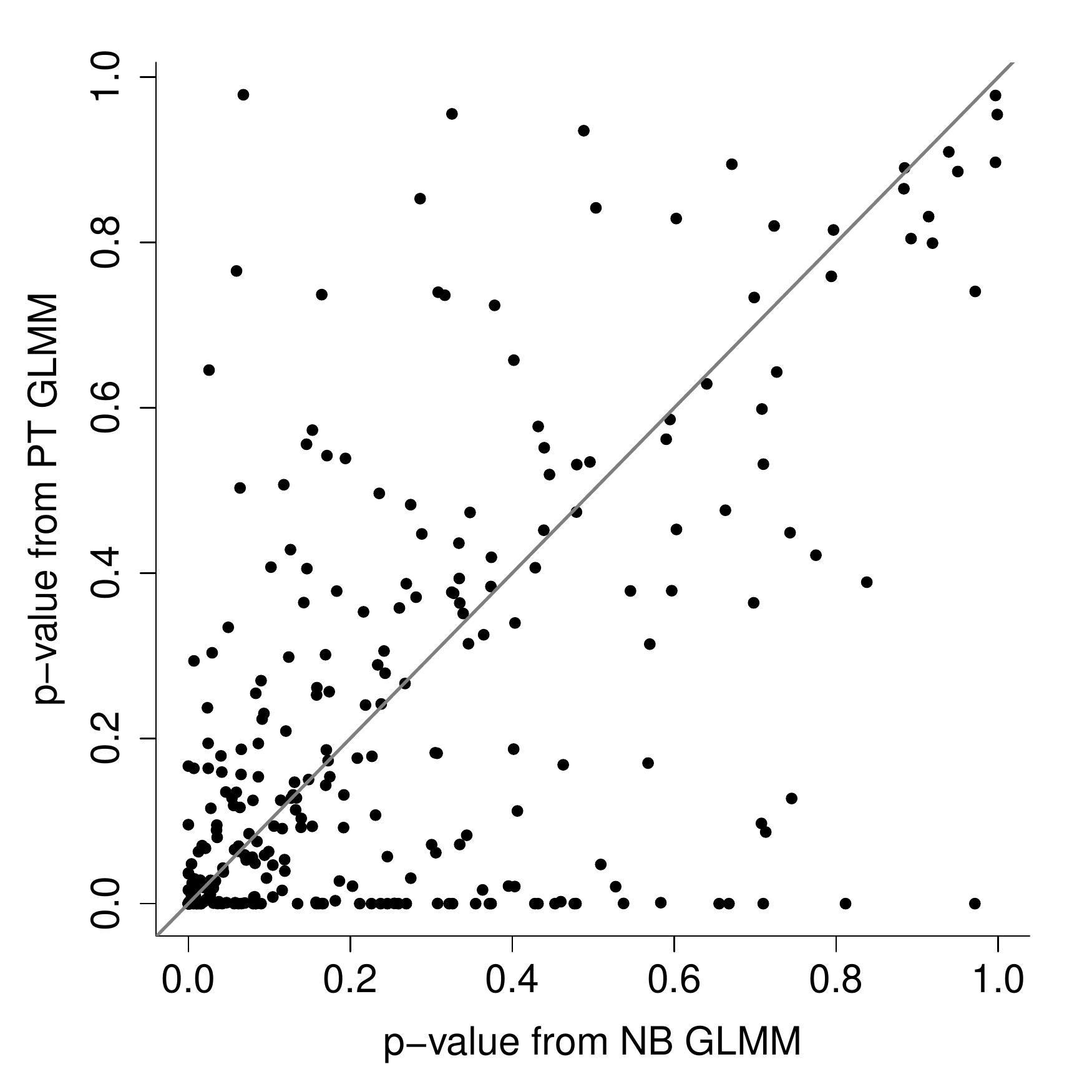}
\caption{Scatter plot comparing the p-values for the hypothesis $H_{01}: \beta_{g1} = \beta_{g3} = 0$ computed using the NB ($x$ axis) and PT ($y$ axis) mixed models.}
\label{fig:appl-p}
\end{figure}

In Figure \ref{fig:appl-p} we compare the p-values for $H_{01}: \beta_{g1} = \beta_{g3} = 0$ obtained from the PT mixed model with those that would have been obtained if we had used a NB mixed model, focusing on the 311 genes for which both models could be fitted. It can be observed that while for several genes the p-values are quite similar, there are also numerous instances of genes for which the choice between NB and PT matters, as it may entail a different conclusion on the significance of the null hypothesis. In particular, p-values obtained from the PT mixed model are smaller than those from the NB mixed model in 178 cases, and larger 133 times; after application of the Benjamini-Hochberg multiple testing correction, 44 genes are found as significant ($\alpha = 0.05$) with both the NB and the PT mixed model; furthermore, 12 genes are identified as significant just by the NB mixed model, whereas 53 genes are found as significant only by the PT mixed model. Overall, these results suggest that while there appears to be a good level of agreement between the two models, identifying for each gene the most suitable distributional shape within the PT family of distributions, rather than arbitrarily fixing $a_g = 0$ and choosing the NB distribution, can have important consequences on the identification of differentially expressed genes and on downstream analyses.

\section{Discussion}
\label{sec5}

High-throughput RNA-sequencing technologies have become the standard method to measure gene expression, progressively replacing microarrays. RNA-seq generates count data that exhibit special features such as overdispersion, zero-inflation and heavy tails. The analysis of RNA-seq data arising from longitudinal experiments should be able to deal with such features and, at the same time, to account for the correlation arising from the availability of repeated measurements from the same individual. However, often such data are analysed with methods that either do not account for overdispersion, zero-inflation and heavy tails \citep{esnaola2013}, or fail to model the dependence between repeated measurements \citep{cui2016}.

In this article we have introduced ptmixed, a GLMM based on the PT distribution that can be employed to analyse of longitudinal sequencing data. On the one hand, the reliance of ptmixed on the PT distribution allows to flexibly model the typical distributional shapes of RNA-seq counts and, at the same time, to perform automatic model selection thanks to the PT power parameter, which allows to distinguish between several overdispersed distributions that can be obtained as special cases of the PT \citep{esnaola2013}. On the other hand, ptmixed properly models the correlation between repeated measurements obtained from the same subject through the specification of subject-specific random effects. The applicability of ptmixed extends beyond RNA-seq data, embracing all scenarios where non-independent measurements of a discrete, overdispersed response variable are available; such settings can arise, for example, in ecology \citep{bolker2009} and in clinical trials \citep{albert1999}.

Our proposal represents the first theoretical development and software implementation of a mixed model based on the PT distribution. An important contribution of our work is to have shown that an accurate evaluation of the integrals in \eqref{likelihood} makes it possible to model longitudinal overdispersed counts using a GLMM that is more flexible than the commonly used Poisson and NB GLMMs. In our experience, we have observed that the rate of success for the estimation of the PT mixed model can be affected by the sample size and on whether the values of $a$ and $\sigma^2$ lie on the boundary of the parameter space; in particular, convergence appears to be easier with larger $n$, and to be more difficult to achieve when the true $a$ is very close to 1 or $\sigma^2$ approaches 0.

The simulations presented in Sections \ref{sub:ptsim} and \ref{sub:agh-lapl} show that ptmixed allows to accurately perform parameter estimation and hypothesis testing, and demonstrate the advantage connected to the use of the AGHQ instead of the LA.
Furthermore, in Section \ref{sub:gwidesim} we presented a comparison between ptmixed and alternative modelling approaches through two genomewide simulations. Those simulations showed that analysing longitudinal sequencing data with methods that do not properly model the repeated measurements design (assuming independence or introducing subject-specific fixed effects) yield inaccurate parameter estimates (Tables \ref{tab:rmse-simC} and \ref{tab:rmse-simD}) and fail to control the type I error in tests on between-subjects effects (Table \ref{tab:fpr-simD}). On the other hand, they showed that use of GLMM approaches such as ptmixed and ShrinkBayes make it possible to accurately estimate and test both within- and between-subject effects.

The results of Section \ref{sub:gwidesim} further allowed to compare ShrinkBayes and ptmixed. As concerns parameter estimation, the RMSEs of ShrinkBayes were found to be 1.5 / 2 times smaller than those of ptmixed for within-subject effects (Table \ref{tab:rmse-simC}), but 5 times larger than those of ptmixed for between-subjects effects (Table \ref{tab:rmse-simD}). Both methods successfully controlled the FDR at $\alpha = 0.05$ in simulation C (Table \ref{tab:fpr-simC}), but ShrinkBayes did not control it when $n = 10$ in simulation D. Interestingly, ptmixed was somewhat more powerful than ShrinkBayes in simulation C (Table \ref{tab:tpr-simC}), while the opposite was found in simulation D (Table \ref{tab:tpr-simD}). Finally, the estimation of ptmixed was faster than ShrinkBayes with small sample sizes, but slower with larger sample sizes. Overall, the comparison pointed out that none of the two methods uniformly outperformed the other, suggesting that both methods are valid modelling approaches with their own advantages and weaknesses.
Aside from those performance measures, there are a few additional features of ptmixed that make it more flexible than ShrinkBayes: as a matter of fact, ptmixed is capable of modelling not only zero-inflation but also heavy-tails, it allows to select a different distributional shape for each gene rather than fitting the same distribution to all genes, and it allows to easily test not only simple univariate hypotheses involving a single parameter, but also more general multivariate hypotheses involving linear combinations of parameters and / or the joint significance of two or more parameters.

In Section \ref{sec4} we illustrated an application of ptmixed to data from a longitudinal RNA-seq experiment involving \textit{WT} and \textit{mdx} mice. Through this application we showed not only that ptmixed is capable to identify as differentially expressed many genes known to play a major role in the evolution of Duchenne Muscular Dystrophy, but also that using the more general PT GLMM instead of the simpler NB GLMM can yield substantially different conclusions on which genes are differentially expressed. This result points out the importance of properly modelling of zero-inflation and heavy-tails not just from a theoretical point of view but also from an applied perspective, since doing so can substantially change the conclusions of real data analyses.

\section*{Supplementary materials}
Supplementary materials for this paper are available from \url{http://www.statmod.org/smij/archive.html}.

\section*{Acknowledgements}
The authors gratefully acknowledge funding from the \textit{Duchenne Parent Project NL} foundation (\url{https://duchenne.nl}) and from the \textit{Association Francaise contre les Myopathies} (grant number 19118).

\bibliographystyle{apa}
\bibliography{refs}

\end{document}